\documentclass[a4paper]{article}

\usepackage{INTERSPEECH2021}
\usepackage{multirow}

\title{Half-Truth: A Partially Fake Audio Detection Dataset}
\name{Jiangyan Yi$^{1,*}$, Ye Bai$^{1,2,*}$\thanks{* denotes equal contribution to this work.}, Jianhua Tao$^{1,2}$, Haoxin Ma$^{1,2}$, Zhengkun Tian$^{1,2}$, Chenglong Wang$^{1,2}$, Tao Wang$^{1,2}$, Ruibo Fu$^{1}$}
\address{
  $^1$National Laboratory of Pattern Recognition, Institute of Automation, Chinese Academy of Sciences \\
$^2$University of Chinese Academy of Sciences, Beijing, China}
\email{\{jiangyan.yi, ye.bai, jhtao, haoxin.ma, zhengkun.tian, chenglong.wang, tao.wang, ruibo.fu\}@nlpr.ia.ac.cn}

\begin{document}

\maketitle
\begin{abstract}
  Diverse promising datasets have been designed to further the development of fake audio detection, such as ASVspoof databases.
  However, previous datasets ignore an attacking situation, in which the hacker hides some small fake clips in real speech audio. This poses a serious threat since that it is difficult to distinguish the small fake clip from the whole speech utterance. Therefore, this paper develops such a dataset for half-truth audio detection (HAD). Partially fake audio in the HAD dataset involves only changing a few words in an utterance. The audio of the words is generated with the very latest state-of-the-art speech synthesis technology. We can not only detect fake uttrances but also localize manipulated regions in a speech using this dataset. Some benchmark results are presented on this dataset. The results show that partially fake audio presents much more challenging than fully fake audio for fake audio detection. The HAD dataset is publicly available\footnote{\url{https://zenodo.org/records/10377492}}.
\end{abstract}

\section{Introduction}

Over the last few years, the technology of speech synthesis has made significant improvement with the development of deep learning \cite{Wang2017Tacotron,Shen2018Wavenet, Wang2018Style}. The models can generate realistic and human-like speech. It is hard for most people to distinguish the generated audio from the real. However, this technology also poses a great threat to the global political economy and social stability if some attackers and criminals misuse it with the intent to cause harm. Therefore, an increasing number of efforts \cite{Wu2015ASVspoof, Kinnunen2017ASVspoof, Todisco2019ASVspoof,Chen2020Gen, Wang2020Deep, Wu2020LCNN} have been made to detect the fake audio recently. A diverse set of databases also have been designed to further the development of this research.

Previously, the majority of the datasets are aimed to detect spoofed utterances for automatic speaker verification systems.
In 2004, Lau et al. \cite{Lau2004Impdataset} have designed an impersonation database for investigating the vulnerability of speaker verification. A small Finnish impersonation database have been created by Hautamaki et al. \cite{Hautamaki2013Impdataset} in 2013.
A few attempts have been made to design individual spoofing databases focused on one speech synthesis \cite{Leon2010Eval, Wang2020ASVspoof} or one voice conversion approach \cite{Bonastre2007Art, Kinnunen2012Vul, Alegre2013A, Kons2013Voice} . Some spoofing databases have been designed to compare with different spoofing methods. A spoofing database have been designed by Wu et al. \cite{Wu2013Vulnerability} involving replay attacks and a simple voice conversion method. Alegre et al. \cite{Alegre2013A} have designed a database including artificial signal spoofing attacks, one voice conversion and speech synthesis approach. Wu et al. \cite{Wu2015SAS} have developed a standard spoofing database SAS including a wide variety of spoofing methods of speech synthesis and voice conversion. Based on the SAS database, the first ASVspoof challenge \cite{Wu2015ASVspoof} have been organized involving only the detection of spoofed speech in 2015. Replay attack is also a key concern among other possible attacks. Therefore, the ASVspoof 2017 corpus including only replay attack is designed for the ASVspoof 2017 challenge \cite{Kinnunen2017ASVspoof}. The ASVspoof 2019 database \cite{Wang2020ASVspoof} consists of synthetic, converted and replayed speech. Previous ASVspoof databases focus on detection of unforseen attack in microphone channel. A spoof dataset in telephone channel is desinged for speaker verification systems by Lavrentyeva et al. \cite{Galina2019Phone}.

\begin{figure}[t]
  \centering
  \includegraphics[width=9.0cm,height=4.0cm]{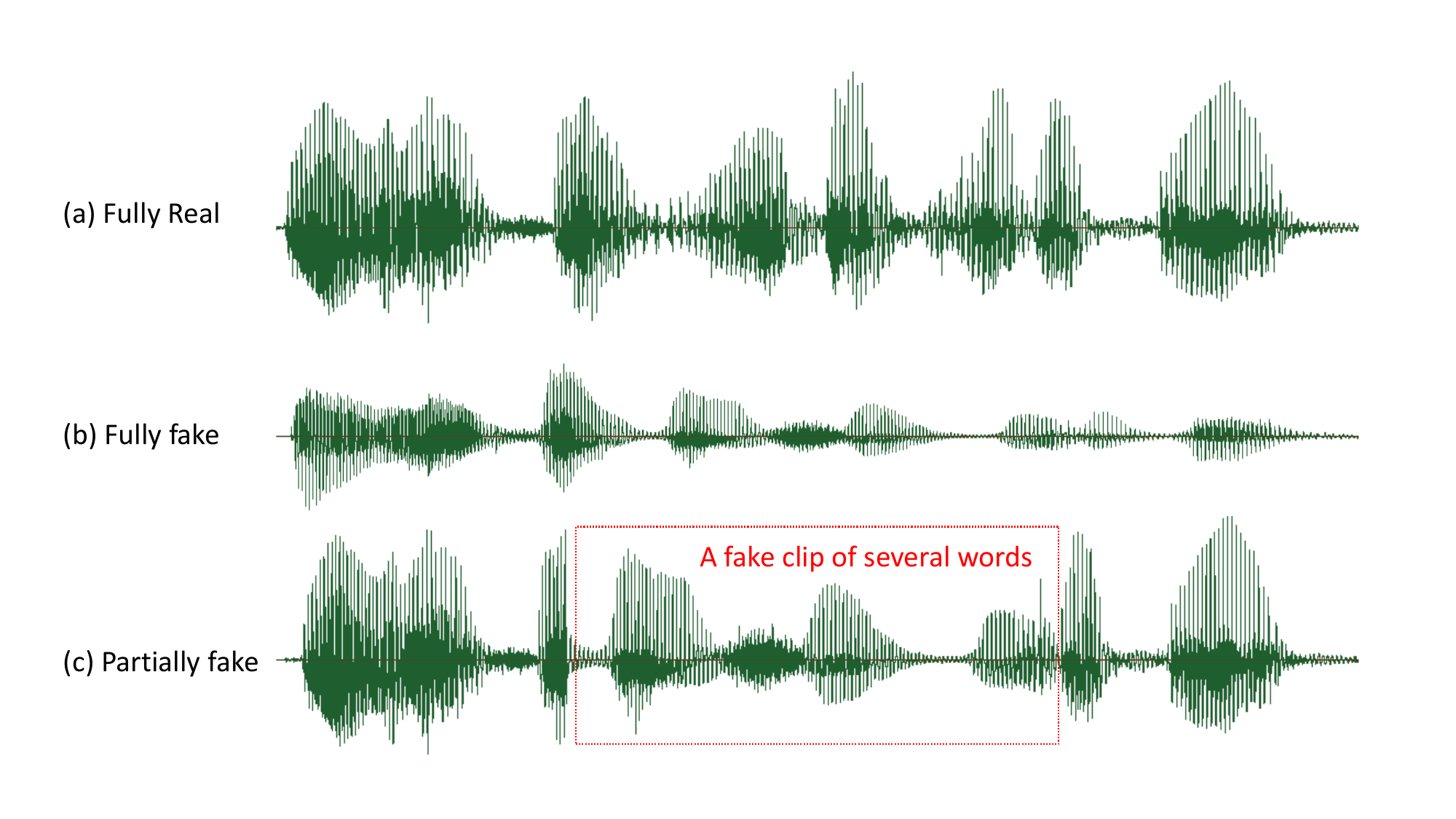}
  \caption{Waveforms of example utterances. (a) illustrates a fully real utterance. (b) shows a fully fake utterance. (c) denotes a partially fake utterance only changing a few words.}
  \label{fig:fake_example}
  \vspace{-10pt}
  \vspace{-10pt}
\end{figure}

Recently, a few of the datasets are designed mainly for fake audio detection systems. A dataset for synthetic speech detection is created by Reimao et al. \cite{Reimao2020For}. The dataset contains the fake speech generated by the open-sourced tools using the latest speech synthesis technology. Wang et al. \cite{Wang2020Deep} have built a English and Mandarin fake dataset with an open-sourced voice conversion and speech synthesis tool.

Previous fake databases are very important for fostering spoofed speech detection research. The ASVspoof databases especially have played a key role in the development of this research.
The fake audio in all previous datasets is fully generated by utterance-level as shown in Figure~\ref{fig:fake_example}~(b). However, previous datasets ignore a fake situation, in which several small fake clips are hided in a real speech audio as shown in Figure~\ref{fig:fake_example}~(c). This poses a serious threat since that it is  not easy to know what changed if attackers and criminals use synthetic audio to change a few words in a speech.

Therefore, this paper reports our progress in developing such a partially fake corpus involving changing a few words in an utterance. The dataset is named Half-truth Audio Detection (HAD). The audio of the words is generated with the very latest state-of-the-art speech synthesis technology, such as global style token (GST) Tacotron \cite{Wang2018Style,Skerry2018Towards}. We describe a preliminary set of benchmark results for detecting fake utterances and localize the manipulated intervals in a speech.

To the best of our knowledge, this is the first attempt to design such a partially fake dataset. The HAD dataset provides the location information of the fake segment of the partially fake audio. Thus, researchers can evaluate the performance of a detection model of localizing the fake part in the partially fake audio.
The HAD corpus will be publicly available soon.

The rest of this paper is organized as follows. Section 2 describes the design policy of the dataset. Evaluation metric is introduced in Section 3. Section 4 presents the experiments and baselines. This paper is concluded in Section 5.

\begin{figure}[t]
	\centering
	\includegraphics[width=6.0cm,height=4.0cm]{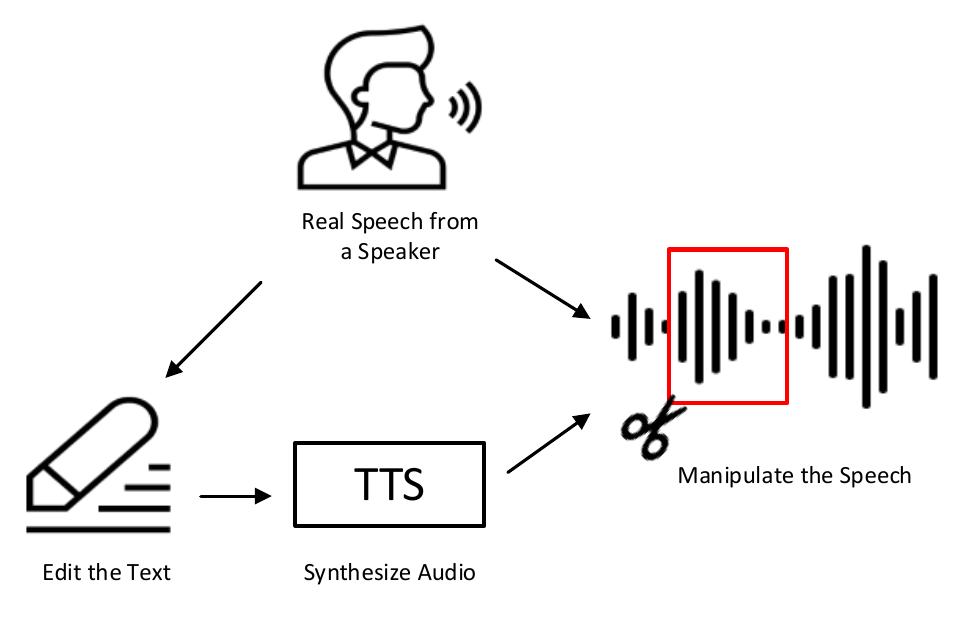}
	\caption{An illustration of the generation of a partially fake audio.}
	\label{fig:policy}
    \vspace{-11pt}
    \vspace{-10pt}
\end{figure}

\section{Dataset design}

Our Half-truth Audio Detection (HAD)  dataset is based on AISHELL-3 corpus \cite{Shi2020aishell3}. It is publicly available\footnote{http://www.openslr.org/93/} and under the Apache license 2.0 . AISHELL-3 is a multi-speaker Mandarin speech corpus for training text-to-speech (TTS) models. The utterances are recorded in a quiet indoor environment using high fidelity microphones at a sampling rate of 44.1 kHz and 16 bits-per-sample. It consists of 88035 utterances about 85 hours of recordings from 218 native Chinese mandarin speakers (175 female, 43 male). The set of 218 speakers is partitioned into two speaker-disjoint sets for training and test.
The training set includes 174 speakers involving 64773 utterances about 60 hours.

\subsection{Design policy}

The HAD dataset is designed to evaluate and analyze the methods of detecting and localizing partially fake audio. Because the synthesized audio is only a small part in the generated fake audio, it is more difficult to distinguish whether the whole utterance is fake or not than purely synthesized audio.
To compare the performance of the detection system, we also provide fully real audio and fully fake audio. Therefore, the dataset consists of two subsets: (1) Partially fake audio set (\textit{Partial}). (2) Fully real and fake audio set (\textit{Full}). Thus, researchers can compare methods using two datasets. Researchers also can evaluate how well a detection system localize the fake part in the partially fake audio.

The core of the HAD dataset is the partially fake audio. The generation of a partially fake audio is illustrated in Figure \ref{fig:policy}. The generation procedure consists of three steps:
(1) Editing content of real speech of a speaker.(2) Synthesizing audio with the edited text. (3) Manipulating the original real speech with the synthesized audio.

\subsection{Editing strategies of content}\label{sec:edit}
\begin{table}
\caption{The statistics of used named entities.}
\vspace{-8pt}
\label{tab:ne}
\centering
\begin{tabular}{cccccc}
 \toprule
\textbf{{Set}} &
{\#\textbf{PER}} &
{\#\textbf{LOC}} &
{\#\textbf{ORG}} &
{\#\textbf{TIME}} &
{\#\textbf{Total}} \\
 \midrule
{Inner} & 4232  & 2731  & 2401  & 1329  & 10693 \\
{Pool} &  5154 & 3307 &  3270 & 1381 & 13112 \\
 \bottomrule
\end{tabular}
    \vspace{-10pt}
\end{table}

\begin{table*}
\caption{The description of the HAD datset which consists of two subsets: Partial and Full.}
\vspace{-8pt}
\label{tab:datasetinfo}
\centering
\begin{tabular}{ccllcccccc}
 \toprule
 \multirow{1}{*}{ Subset } &
\multirow{1}{*}{ Audio type } &
\multirow{1}{*}{ Partition} &
\multirow{1}{*}{\#Total} &
\multirow{1}{*}{\#Female } &
\multirow{1}{*}{\#Male} &
\multirow{1}{*}{\#Southen} &
\multirow{1}{*}{\#Northern} &
\multirow{1}{*}{\#Other } \\
 \hline
 \multirow{4}{*}{\textit{Partial}} & \multirow{4}{*}{Partially fake} & Train & 26554 & 20962 & 5592&6874 & 19330 & 350   \\
 & & Dev. & 8914 & 7388 &1526& 2174 & 6740& 0   \\
 & & Test & 9072 & 7366 &1706& 3490 & 5242 & 340   \\
 &  & Unseen Test & 9072 &7366 &1706&  3490 & 5242 & 340 \\
 \hline
  \multirow{8}{*}{\textit{Full}} & \multirow{4}{*}{Fully fake} & Train & 26554 & 20962 &5592& 6874 & 19330 & 350 \\
 & & Dev. & 8914 & 7388 &1526& 2174 & 6740& 0  \\
 & & Test & 9072 & 7366 &1706& 3490 & 5242 & 340 \\
 &  & Unseen Test & 9072 & 7366 &1706& 3490 & 5242 & 340 \\
 & \multirow{4}{*}{Fully real} & Train & 26554 & 21582 &4972& 8446 & 17631 & 477 \\
 & & Dev. & 8914 & 7684 &1230& 3398 & 5516 & 0 \\
 & & Test & 9072 &  7249 &1823& 3870 & 4756 & 446 \\
 &  & Unseen Test & 9072 & 7249 &1823& 3870 & 4756 & 446 \\
 \bottomrule
\end{tabular}
    \vspace{-10pt}
\end{table*}

We edit the text of the real speech to synthesize utterances which have different meaning of the original utterances.
The selecting strategy is by the content of the utterances. The transcription of these utterances including keywords, which influence the meaning of the utterance if they are manipulated.
We mainly use two strategies: (1) Randomly replacing named entities in the text of a utterance. (2) Replacing a word which reflects the attitude with the corresponding antonym. For simplicity, each edited sentence only contains one replacing operation.

We use open source toolkit jieba\footnote{https://github.com/fxsjy/jieba} for word segmentation and named entity recognition. There are four kinds of entities: person (PER), location (LOC), organization (ORG) and time (TIME).
``Total'' denotes all the entities. The information of the used entities is shown in Table \ref{tab:ne}. \#Inner denotes the number of the entities in the content of the real speech. \#Pool means the size of the entity pool which we use to randomly select an entity to replace the corresponding entity in the sentence. For antonym edition, we use 181 word/antonym pairs to change the meaning of a sentence to the contrast.

\subsection{Fully fake audio generation} \label{sec:fullyfake}

We employ a commonly used open source multi-speaker end-to-end TTS code\footnote{https://github.com/syang1993/gst-tacotron.git} to train a speech synthesis model. The model is based on GST \cite{Wang2018Style} and prosody transfer \cite{Skerry2018Towards}. It is capable of generating an audio spectrogram for the style (e.g., tone, pitch) of the target speaker. Then the spectrogram is used to generate audio with a vocoder. We use a high-performance neural vocoder LPCNet \cite{Valin2019lpcnet} with 22-dimensional acoustic features to generate the audio \footnote{https://github.com/mozilla/LPCNet}. The components of the whole TTS system are all open source. Thus, the reproducibility is ensured.

The model is trained using the training set of the AISHELL-3 corpus. The fully fake audio is generated with the model using the original text and the above-mentioned edited text. The fake segment is cut out from the full synthetic utterance. We do not directly synthesize the keyword audio. Because for end-to-end TTS systems, the generation of a whole utterance is natural than one word or phrase.

\subsection{Manipulation for partially fake audio generation}  \label{sec:manipulation}

We manipulate the original real audio with the fully synthetic audio to generate partially fake audio. The manipulating process is as follow. (1) Selecting a keyword in the transcription of the real utterance using the method described in Section~\ref{sec:edit}. (2) Selecting the corresponding synthetic audio generated with the edited content. (3) Normalizing the volume of the audio. (4) Replacing the segment of the selected keyword in the original real utterance with the segment of the manipulated content in the synthetic audio.

We use a well-trained speech recognition system to generate timestamps of each character of the utterances by forced alignment. Thus, we can replace the segment of the selected keyword automatically. We use pydub library\footnote{https://github.com/jiaaro/pydub} for manipulation. Because we use different methods to manipulate these utterances, we can generate more partially fake utterances than the corresponding real speech.

\subsection{Dataset composition}

There are two subsets in the HAD dataset: \textit{Partial} and \textit{Full}.
The real utterances of the \textit{Full} set are selected from the training set of AISHELL-3. The fake utterances of the \textit{Full} set are generated by the TTS model trained using the training set of AISHELL-3  described  Section~\ref{sec:fullyfake}.  The partially fake utterances of the \textit{Partial} set are generated by manipulating the real speech using the synthesized audio described in  Section~\ref{sec:manipulation}.

The training set (Train), development (Dev.) set, and test set (Test) are partitioned with 6:2:2. There are no overlaps among the speakers of the sets. We make the total number of utterances of each subset for different parts be equal so that the performance can be compared fairly. To test the generalization for accents, the data contains different regional accents: southern accent (Southern), northern accent (Northern) and other accent (Other). In order to evaluate generalization of the models, we add an unseen test set (Unseen Test) both in \textit{Partial} and \textit{Full}. The speech data of unseen test sets is generated by the other TTS model using the improved LPCNet vocoder with 32-dimensional acoustic features \cite{2014Comparative}.
Table~\ref{tab:datasetinfo} shows the description of the HAD dataset.

\section{Evaluation metric}
\begin{figure}[t]
	\centering
	\includegraphics[width=1.0\linewidth]{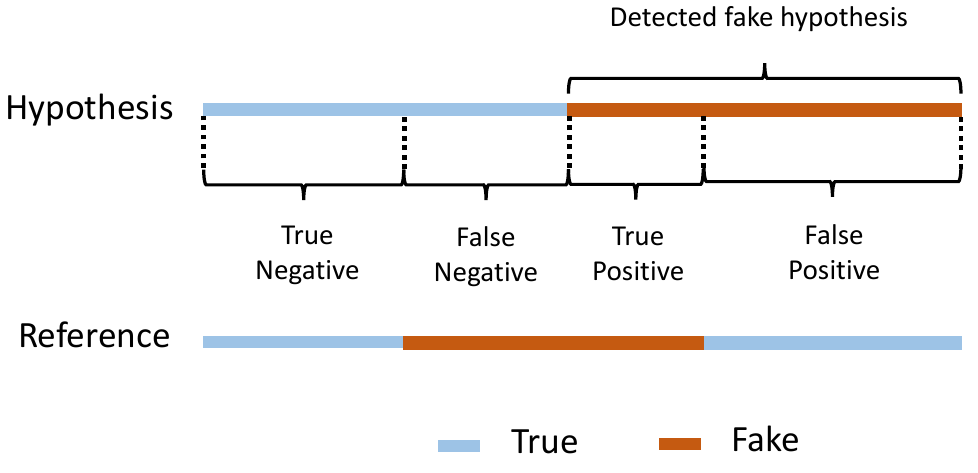}
	\caption{An illustration of the segment-level metric.}
	\label{fig:eval}
    \vspace{-10pt}
    \vspace{-10pt}
\end{figure}

In this paper, we not only provide the results of the whole utterance, but also localize manipulated audio regions. Therefore, we use two kinds of metrics described in the following.

\textbf{Utterance-level evaluation.} The models of utterance-level binary classification are evaluated in terms of equal error rate (EER). More details of metrics can be found in \cite{Wu2015ASVspoof}.

\textbf{Segment-level evaluation.} We also localize manipulated audio regions. The illustration of this evaluation is shown in Figure \ref{fig:eval}. We evaluate the overall segment-level fake audio detection performance, which includes the performance of the localization of the model, with standard precision ($P$), recall ($R$) and $F_1$-score ($F_1$). These metrics are based on the duration of each segment. The computation is as follow.

\vspace{-10pt}
\begin{align}\label{attention}
P &=\frac{TP}{TP+FN}  \\
R &=\frac{FN}{TP+FP}  \\
F_1 &=\frac{2 P R}{P+R}
\end{align}
\vspace{-10pt}

 $TP$ means the duration of the overlap between the detected fake hypothesis and the reference of the fake segment. $FP$ denotes the duration of the overlap between the detected fake hypothesis and the wrongly detected fake segment. $FN$ denotes the duration of the detected fake hypothesis and the undetected fake segment.

\section{Initial benchmarking experiments}

A series of benchmark experiments are conducted on the HAD dataset.

\subsection{Experimental setup}

Motivated by the ASVspoof challenge \cite{Todisco2019ASVspoof}, we use Gaussian mixture model (GMM) and light convolutional neural network (LCNN) \cite{Wu2019lcnn} to train baseline models.
We follow the officially released implementation toolkit~\footnote{https://www.asvspoof.org/asvspoof2019} by ASVspoof 2019 to extract features and build GMM based classifiers. LCNN based classifiers are implemented with the Pytorch toolkit~\footnote{https://github.com/pytorch/pytorch}.

Constant Q cepstral coefficients (CQCC) \cite{2017Constant} are used as the input features for all models. The features are extracted with a 20-ms sliding window with a 10-ms shift. The CQCC features include 29 CQCC coefficients appended by energy (C0 or $0^{th}$ cepstral coefficients) and 7 different combinations of static, delta and acceleration parameters.

The GMM based classifiers have 512-component. They are trained with an expectation maximisation (EM) algorithm with random initialisation. EM is performed until likelihoods converge on the development set.
The basic architecture of the LCNN model is same to that in \cite{Lavrentyeva2017Audio}.
The LCNN model consists of five convolution layers, ten Max-Feature-Map layers and two fully connected layers. More details of the description of the LCNN are found in \cite{Lavrentyeva2017Audio}. The classifiers of GMM and LCNN are standard 2-class discriminators. The classes are real and fake. The GMMs and LCNN models are trained with the real and fake audio utterances in the training set.
The development sets are utilized for selecting models and hyper parameters. The training terminates when a little improvement
between two epochs on the development set has been observed.

\begin{table}[t]
\vspace{-10pt}
	\caption{EERs(\%) of models trained using fully fake and fully real data on test sets.}
\vspace{-8pt}
	\label{tab:fullyeer}
	\centering
	\begin{tabular}{cccccc}
		\toprule
		\multirow{2}{*}{Model} & \multirow{2}{*}{Dev.} & \multicolumn{2}{c}{ Test} & \multicolumn{2}{c}{Unseen Test} \\ \cmidrule(l){3-6}
		&                       & \textit{Full}       & \textit{Partial}     & \textit{Full}      & \textit{Partial}      \\ \midrule
		GMM                    & 0.022                 & 0.022      & 33.38       & 0.52      & 35.13        \\
		LCNN                   & 0.28                  & 0.11       & 38.12       & 0.61      & 39.52        \\ \bottomrule
	\end{tabular}
\end{table}

\begin{table}[t]
	\caption{EERs(\%) of models trained using partially fake data on test sets.}
\vspace{-8pt}
	\label{tab:partiallyeer}
	\centering
	\begin{tabular}{cccccc}
		\toprule
		\multirow{2}{*}{Model} & \multirow{2}{*}{Dev.} & \multicolumn{2}{c}{ Test} & \multicolumn{2}{c}{Unseen Test} \\ \cmidrule(l){3-6}
		&                       & \textit{Full}      & \textit{Partial}      & \textit{Full}      & \textit{Partial}      \\ \midrule
		GMM                    & 10.65                 & 0.14      & 12.67        & 0.75      & 17.66        \\
		LCNN                   & 4.57                  & 0.65      & 4.50         & 2.26      & 9.62         \\ \bottomrule
	\end{tabular}
\vspace{-10pt}
\end{table}

\begin{table}[t]
\vspace{-10pt}
	\caption{EERs(\%) of models trained using fully fake and real data and partially fake data on test sets.}
\vspace{-8pt}
	\label{tab:alleer}
	\centering
	\begin{tabular}{cccccc}
		\toprule
		\multirow{2}{*}{Model} & \multirow{2}{*}{Dev.} & \multicolumn{2}{c}{ Test} & \multicolumn{2}{c}{Unseen Test} \\ \cmidrule(l){3-6}
		&                       & \textit{Full}      & \textit{Partial}      & \textit{Full}      & \textit{Partial}      \\ \midrule
		GMM                    & 7.54                  & 0.04      & 13.92        & 0.43      & 19.22        \\
		LCNN                   & 1.74                  & 0.54      & 2.90         & 2.11      & 6.99         \\ \bottomrule
	\end{tabular}
\end{table}

\begin{table}[t]
	\caption{$P$(\%),$R$(\%) and $F_1$(\%) of models trained using partially fake data on partially fake test sets.}
\vspace{-8pt}
	\label{tab:partiallyf1}
	\centering
	\begin{tabular}{ccccccc}
		\toprule
		\multirow{2}{*}{Model} & \multicolumn{3}{c}{Test} & \multicolumn{3}{c}{Unseen Test} \\ \cmidrule(l){2-7}
		& $P$ & $R$ & $F_1$ & $P$ & $R$ & $F_1$        \\ \midrule
		GMM                    & 36.69    & 84.86    & 45.71   & 34.52     & 83.78    & 45.19    \\
		LCNN                   & 78.88    & 97.16    & 87.07   & 76.36     & 96.85    & 88.26    \\ \bottomrule
	\end{tabular}

\end{table}

\begin{table}[!htb]
	\caption{$P$(\%),$R$(\%) and $F_1$(\%) of models trained using fully fake and real data and partially fake data on test sets.}
\vspace{-8pt}
	\label{tab:allseenf1}
	\centering
	\begin{tabular}{cccccc}
		\toprule
		&    & \multicolumn{2}{c}{ Test} & \multicolumn{2}{c}{Unseen Test} \\ \cmidrule(l){3-6}
		&    & GMM         & LCNN       & GMM          & LCNN        \\ \midrule
		\multirow{3}{*}{\textit{Full}}    & $P$  & 60.32       & 75.18      & 61.51        & 74.70       \\
		& $R$  & 99.80       & 86.64      & 99.76        & 85.25       \\
		& $F_1$ & 75.25       & 83.17      & 74.05        & 82.08       \\ \midrule
		\multirow{3}{*}{\textit{Partial}} & $P$  & 43.17       & 71.62      & 41.20        & 72.54       \\
		& $R$  & 87.52       & 93.86      & 85.32        & 91.88       \\
		& $F_1$ & 56.37       & 82.15      & 54.87        & 82.66       \\
		\bottomrule
	\end{tabular}
\vspace{-15pt}
\end{table}

\subsection{Baselines: utterance-level classification}
In the first group of experiments, we report the class of the whole utterance for each audio.
The results are reported on seen and unseen test sets of HAD datasets. Each kind of test set includes two test sets from the \textit{Partial} and \textit{Full} set. The models are evaluated in terms of EER on four test sets. The results of the models trained using fully fake and fully real data are listed in Table \ref{tab:fullyeer}. The results of trained using partially fake data are reported in Table \ref{tab:partiallyeer}. Table \ref{tab:alleer} reports the results of trained using partially fake and fully fake and real data.

The results show that although a detection system trained on fully fake and real data achieves a good performance, its performance on the partially fake test set degrades significantly. These results also show that even a system is trained on partially fake data, its performance is still poor on the partially fake test set. This group of experiments show that it is challenging to detect the fake audio from partially fake data.

\subsection{Baselines: segment-level classification}

In the second group of experiments, we predict the class of every segment in an utterance. The classifiers of GMM and LCNN are frame-level standard 2-class discriminators.
The architecture of LCNN based frame-level discriminators is only without pooling layers compared to that of LCNN based utterance-level LCNN classifiers.

We employ average smoothing strategy in test stage to convert the results from frame-level to segment-level. We take the average of the scores of adjacent 5 seconds as the final score. The decision thresholds of the systems are set to 0.5.

The results are reported on seen and unseen test sets of HAD datasets. The models are evaluated by segment-level in terms of $P$,  $R$ and $F_1$ on test sets. The results of systems trained with partially fake data are reported in Table \ref{tab:partiallyf1}. Table \ref{tab:allseenf1} report the results of trained using all the partially fake and fully fake and real data on on seen and unseen test sets. These provide evaluation results of localization performance of a detection system for partially fake data. It is shown that it is challenging to localize the manipulated regions in partially fake audio.

\section{Conclusions}

This paper describes the design policy, fake data generation, manipulation and metrics of the HAD dataset. This is the
first dataset that considers a half-truth audio that only contains a fake segment of a few words. The dataset has two specific subsets: partially fake, fully fake and real audio. The fake speech is generated using the state-of-the-art end-to-end acoustic model and the neural vocoder. This paper also reported the benchmark results on this dataset. The results show that it is more difficult to detect the partially fake audio than the fully fake audio.
This database provides location information of the fake segment, frame-level and utterance-level ground truth labels. We strongly believe that the HAD dataset will further accelerate and foster research on fake audio detection and media forensics. Future work includes involving more kinds of fake types and developing datasets for other languages.

\section{Acknowledgements}

This work is supported by the National Key Research and Development  Plan of China (No.2020AAA0140003), the National Natural Science Foundation of China (NSFC) (No.61831022, No.61901473, No.61771472, No.61773379), and Inria-CAS Joint Research Project (No.173211KYSB20190049).

\bibliographystyle{IEEEtran}

\bibliography{mybib}


\end{document}